# Transverse Electron Focusing in Systems with Spin-Orbit Coupling

Gonzalo Usaj and C. A. Balseiro
*Instituto Balseiro and Centro Atómico Bariloche, Comisión Nacional de Energia Atómica, 8400 San Carlos de Bariloche, Argentina.*

We study the transverse electron focusing in a two dimensional electron gas with Rashba spin-orbit coupling. We show that the interplay between the external magnetic field and the spin-orbit coupling gives two branches of states with different cyclotron radius within the same energy window. This effect generates a splitting of the first focusing peak in two contributions. Each one of these contributions is spin polarized. The surface reflection mixes the two branches and the second focusing peak does not present the same effect. While for GaAs/AlGaAs heterostructures the effect is small, in systems like InSb/InAlSb the effect should be clearly observed.

Transverse electron focusing in a two dimensional electron gas (2DEG) was reported almost fifteen years ago.[1] Since then, several experiments have probed electron focusing in different semiconducting heterostructures and geometries. An important feature of these experiments is that electron transport is in the ballistic regime despite typical distances between emitter and collector are of the order of $1\mu m$. This is due to the long mean free path ($\sim 40\mu m$) of the 2DEG. Very recently, this technique was used to collect electrons coming from a point contact[2] or quantum dot[3,4] into another point contact acting as a voltage probe. Additionally, a large in-plane magnetic field was applied in order to have a spin-dependent transmission through both the emitter and the collector. This opened the possibility to inject and detect spin polarized electrons in a 2DEG without using ferromagnetic materials. Producing and detecting spin polarized currents, or pure spin currents,[5] is the ultimate goal of spintronics.[6] In this context, the spin-orbit interaction in 2DEGs has generated great interest since this intrinsic effect could also be used to manipulate and control the electron's spin.[7] For this reason, during the last years a substantial amount of work has been devoted to study its effect on the transport properties of nanostructures and 2DEGs.[8–16]

In this work we present results for transverse electron focusing in systems with spin-orbit coupling. Since the focusing effect is dominated by edge states,[17] our results concern the structure and nature of edge states in a 2DEG with spin orbit coupling.[18] Our starting point is a 2DEG with Rashba spin-orbit coupling[19] and an external magnetic field $B_z$ perpendicular to the plane containing the electron gas

$$H=\frac{1}{2m^*}(P_x^2+P_y^2)+\frac{\alpha}{\hbar}(P_y\sigma_x-P_x\sigma_y)-\frac{1}{2}g\mu_B\sigma_z B_z\,, \quad (1)$$

where $P_\eta = p_\eta - (e/c)A_\eta$ with $p_\eta$ and $A_\eta$ being the $\eta$-component of the momentum and vector potential respectively, $\alpha$ the Rashba coupling parameter, $g$ the effective g-factor and $\{\sigma_\eta\}$ the Pauli matrices. For a numerical evaluation of the Green functions of a system with arbitrary shape it is convenient to discretize the space and reduce the model to the following tight-binding Hamiltonian $H=H_0+H_R$ with

$$H_0=\sum_{n,\sigma}\varepsilon_\sigma c^\dagger_{n\sigma}c_{n\sigma} - \sum_{<n,m>,\sigma} t_{nm}\, c^\dagger_{n\sigma}c_{m\sigma}+h.c.\,, \quad (2)$$

where $c^\dagger_{n\sigma}$ creates an electron at site $n$ with spin $\sigma$ and energy $\varepsilon_\sigma = 4t-\sigma g\mu_B B_z/2$, $t=\hbar^2/2m^*a_0^2$ and $a_0$ is the effective lattice parameter—in what follows we use $a_0=5nm$ which is small compared with a typical Fermi wavelength. The summation is carried out on a square lattice and $n=n_x\hat{\boldsymbol{x}}+n_y\hat{\boldsymbol{y}}$ where $\hat{\boldsymbol{x}}$ and $\hat{\boldsymbol{y}}$ are unit vectors in the $x$ and $y$ directions respectively. The hopping matrix element $t_{nm}$ connects nearest neighbors and includes the effect of the diamagnetic coupling through the Peierls substitution.[20] We use the Landau gauge for which $t_{n(n+\hat{\boldsymbol{x}})} = t\exp\left(-\mathrm{i}n_y 2\pi\phi/\phi_0\right)$ and $t_{n(n+\hat{\boldsymbol{y}})} = t$. $\phi=a_0^2 B_z$ is the magnetic flux per plaquete and $\phi_0 = hc/e$ is the flux quantum. The second term of the Hamiltonian describes the Rashba coupling

$$\begin{aligned} H_R = -\lambda\sum_n \Big\{ &\mathrm{i}\left(c^\dagger_{n\uparrow}c_{(n+\hat{\boldsymbol{y}})\downarrow}+c^\dagger_{n\downarrow}c_{(n+\hat{\boldsymbol{y}})\uparrow}\right) \\ &-e^{-\mathrm{i}n_y 2\pi\phi/\phi_0}\left(c^\dagger_{n\uparrow}c_{(n+\hat{\boldsymbol{x}})\downarrow}-c^\dagger_{n\downarrow}c_{(n+\hat{\boldsymbol{x}})\uparrow}\right)\Big\}+h.c.\,, \end{aligned} \quad (3)$$

where $\lambda=\alpha/2a_0$. The band structure and the Fermi surfaces in the absence of the transverse field are shown in Figure 1. The spin-orbit coupling removes the spin degeneracy leading to two bands, $\varepsilon_k = \varepsilon^0_k \pm \Lambda_k$ with $\varepsilon^0_k = 4t-2t(\cos k_x a_0 + \cos k_y a_0)$ and $\Lambda_k = 2\lambda(\sin^2 k_x a_0+\sin^2 k_y a_0)^{\frac{1}{2}}$. Note that if the Fermi energy ($E_{\mathrm{F}}$) is greater than zero—we assume that in what follows—the two bands contribute to the Fermi surface and have the same Fermi velocity.

Let us now consider a system with the geometry shown in Figure 1c. It consists of a semi-infinite 2DEG with two lateral contacts, numbered 1 and 2, at a distance $R$ from each other. Each contact is a narrow stripe with a width of $N_0$ sites and, for simplicity, no spin orbit coupling. They represent point contacts gated to have a single active channel with a conductance $2e^2/h$. Then, they can be approximated by a linear chain as shown in the inset of Figure 1c. The hopping matrix elements coupling the chain with the semi-infinite 2DEG are modulated by the wave function of the point contact transverse mode. Typical experimental setups include also one or two ohmic contacts at the bulk of the 2DEG. They are used to inject currents and measure voltages. The focusing experiment consists of injecting a current $I$ through contact 1 and measuring the voltage $V_2$ generated in contact 2. According to the Landauer-Büttiker formalism for linear response,[21] $V_2/I$ can be written in terms of the conductances $G_{NM}$ between contacts $N$ and $M$. Details of different configurations with three and four contacts have been analyzed in Ref. [1]. The main features of the magnetic field dependence of the focusing peaks are contained in the conductance $G_{12}$ between the two lateral contacts.[17] Therefore, in what follows we present results for the conductance $G_{12}$ in the weak field limit, the relevant regime for focusing experiments, were Landau quantization is not important. The conductance is given by[22,23]

$$G_{12} = \frac{e^2}{h}\mathrm{Tr}\{\Gamma^{(1)}\mathcal{G}^R\Gamma^{(2)}\mathcal{G}^A\}\,, \qquad (4)$$

where $\mathcal{G}^R$ and $\mathcal{G}^A$ are the retarded and advanced Green function matrices respectively and $\Gamma^{(N)}$ is the "coupling matrix" to contact $N$. The matrix elements $\mathcal{G}_{ij}$ of the Green functions are the propagators from site $i$ to site $j$ and $\Gamma^{(N)}_{ij} = \mathrm{i}[\Sigma^R_N - \Sigma^A_N]_{ij}$ with $\Sigma^R_N$ and $\Sigma^A_N$ the self-energies of the retarded and advanced propagators due to the contact $N$. All these quantities are evaluated at $E_{\rm F}$. We first calculate the propagators of the system without the contacts by Fourier transforming in the $x$-direction and generating a continuous fraction for each $k_x$. Having these propagators, the self energies due to the contacts can be easily included using the Dyson equation.[20]

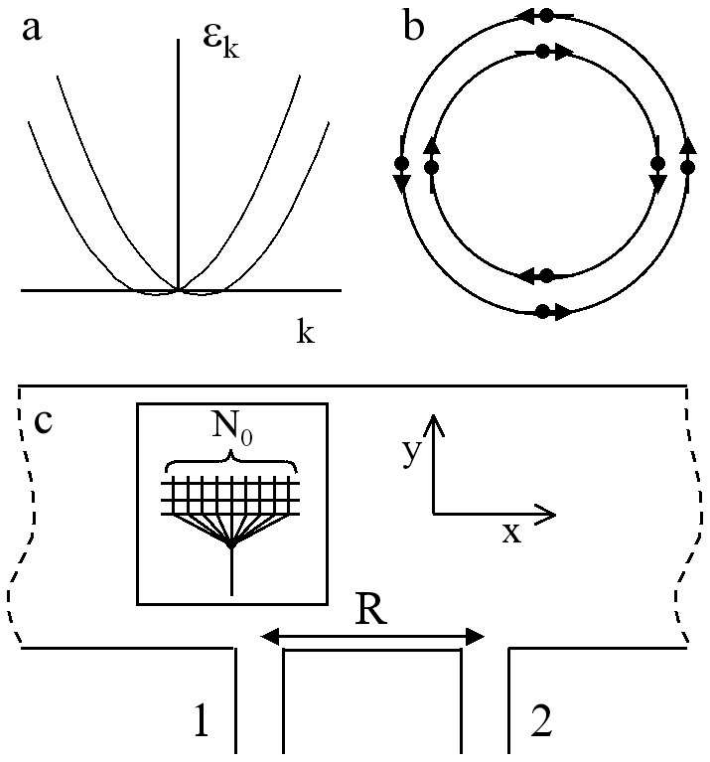

FIG. 1: a) Band structure for an infinite 2DEG with Rashba spin-orbit coupling; b)corresponding Fermi surfaces for $E_{\rm F} > 0$. Note that the eigenstates have the spin in the plane of the 2DEG and perpendicular to the momentum; c) Geometry used to study the transverse focusing. The inset shows a detail of the model used to describe the contacts.

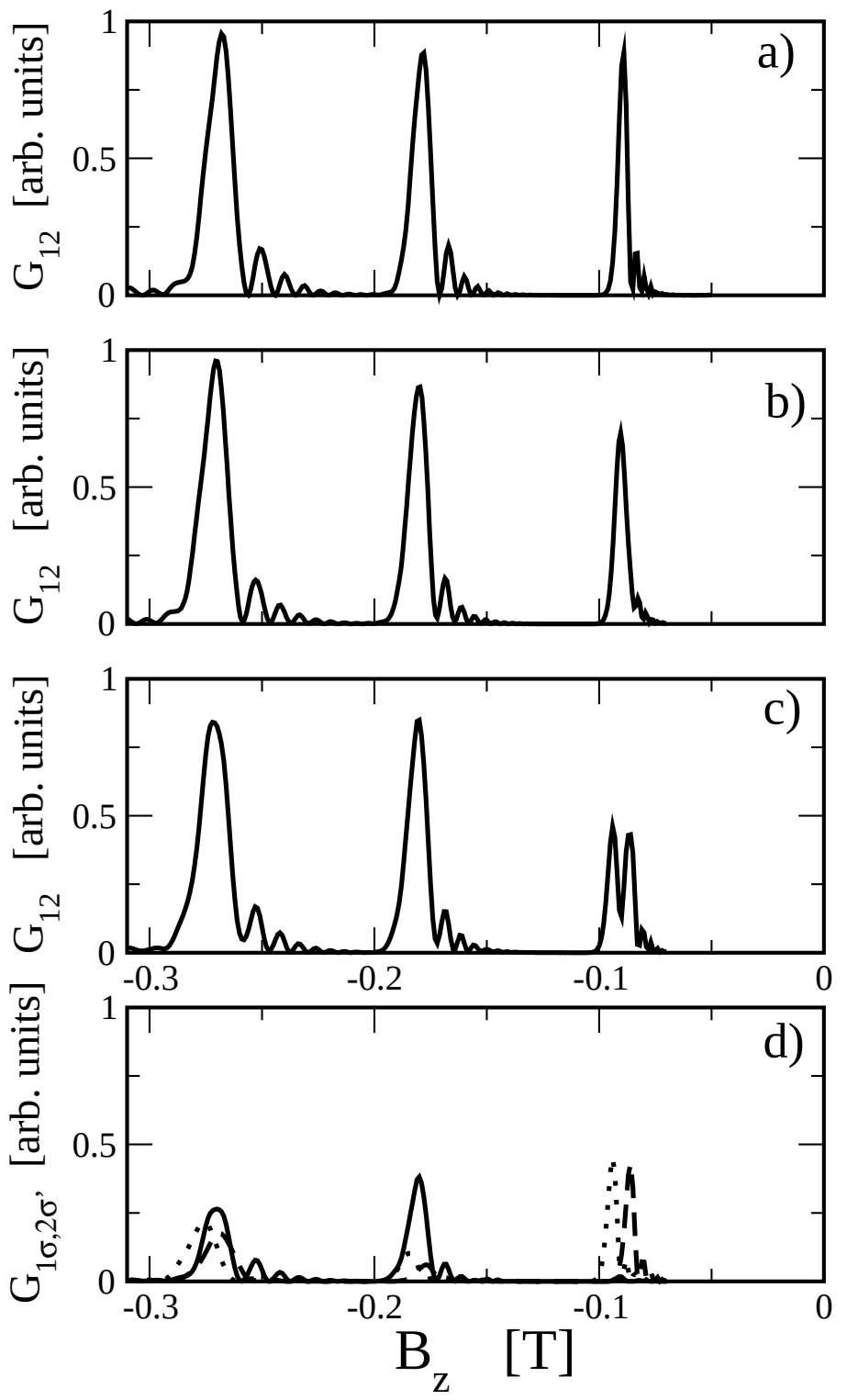

FIG. 2: Conductance $G_{12}$ as a function of the magnetic field $B_z$. The parameters are $E_{\rm F} = 6\,meV$, $R = 1.5\,\mu m$, $N_0 = 13$ and: a) $\alpha = 0$, b) $\alpha = 2\,meVnm$, and c) $\alpha = 5\,meVnm$. In (d) the partial contributions $G_{1\sigma,2\sigma'}$ for the parameters in (c) are shown: $G_{1+,2-}$ dashed line, $G_{1-,2+}$ dotted line and $G_{1+,2+}$ solid line—in this scale $G_{1-,2-}$ can not be distinguished from $G_{1+,2+}$. The spin quantization axis is $x$. The peak's splitting is given by $\Delta B_z = 4(\alpha/R)m^*c/\hbar e \simeq 7mT$.

Figure 2a shows the conductance $G_{12}$ as function of the magnetic field $B_z$ for systems with an intermediate contact width ($N_0 = 13$) and no spin-orbit coupling ($\alpha = 0$). The parameters correspond approximately to GaAs/AlGaAs with an electron density $n_e \simeq 2 \times 10^{11}/cm^2$ and a contact-contact distance $R = 1.5\mu m$ (as used in Ref. [2]). All focusing peaks are well defined and a diffraction-like structure around each peak is obtained. As the contacts width increases the peaks get broader and these structures disappear. We discuss this point in more detail below. In the same figure we also present results for a system having intermediate (figure 2b) and large (figure 2c) spin-orbit coupling. As the spin-orbit parameter $\alpha$ increases, the first focusing peak splits in two well defined contributions. Each contribution is spin-polarized in the $x$-direction. To a good accuracy, the splitting is given by $\Delta B_z = 4(\alpha/R)m^*c/\hbar e$ (cf. Eq. (8) below). Quite remarkably, the second focusing peak does not split. To analyze the spin dependence we separate the total conductance $G_{12}$ in different contributions

$$G_{12} = \sum_{\sigma\sigma'} G_{1\sigma,2\sigma'}\,, \qquad (5)$$

where $G_{1\sigma,2\sigma'}$ represents the conductance from contact 1 to contact 2 with the injected and collected electrons having spin $\sigma$ and $\sigma'$ respectively. In what follows the spin quantization axis is taken in the $x$-direction. Thus, $G_{1+,2-}$ refers to the conductance of electrons injected with spin $\sigma_x = +1$ and collected with spin $\sigma_x = -1$. The partial conductances are plotted in figure 2d. Around the first focusing peak only $G_{+,2-}$ and $G_{1-,2+}$ are large and each one of these contributions leads to a single peak. This is consistent with a semiclassical picture of the spin-orbit interaction: as the momentum is reversed by the action of the external field, the spin rotates from $\sigma_x$ to $-\sigma_x$. In the second focusing peak $G_{1+,2+}$ and $G_{1-,2-}$ dominate; the spin is reversed twice and the electron arrives at contact 2 with the same spin projection $\sigma_x$ it had at the injection point. These results suggest that the splitting of the first focusing peak is due to the existence of two semiclassical cyclotron radii originated in the two spin-orbit bands. At first sight this is surprising since in the absence of the magnetic field the two bands have the same velocity. However, the eigenstates of the 2DEG with a magnetic field and spin-orbit coupling show that for a given energy there are two semiclassical radii. In the infinite system, the spectrum consists of two branches that, for large Landau index $s$, have energies given by[19,24]

$$E_s^\pm \simeq \hbar\omega_c s \pm \frac{\alpha}{l_c}\sqrt{2s}\,, \tag{6}$$

where $\omega_c = eB_z/m^*c$ and $l_c = (\hbar/m^*\omega_c)^{\frac{1}{2}}$. In the Landau gauge, the corresponding large $s$ eigenfunctions are given by

$$\Psi_s^\pm(x,y) \simeq \frac{1}{\sqrt{2L_x}} e^{\mathrm{i}k_x x} \begin{pmatrix} \phi_{s-1}(y-y_c) \\ \pm\mathrm{i}\phi_s(y-y_c) \end{pmatrix} . \tag{7}$$

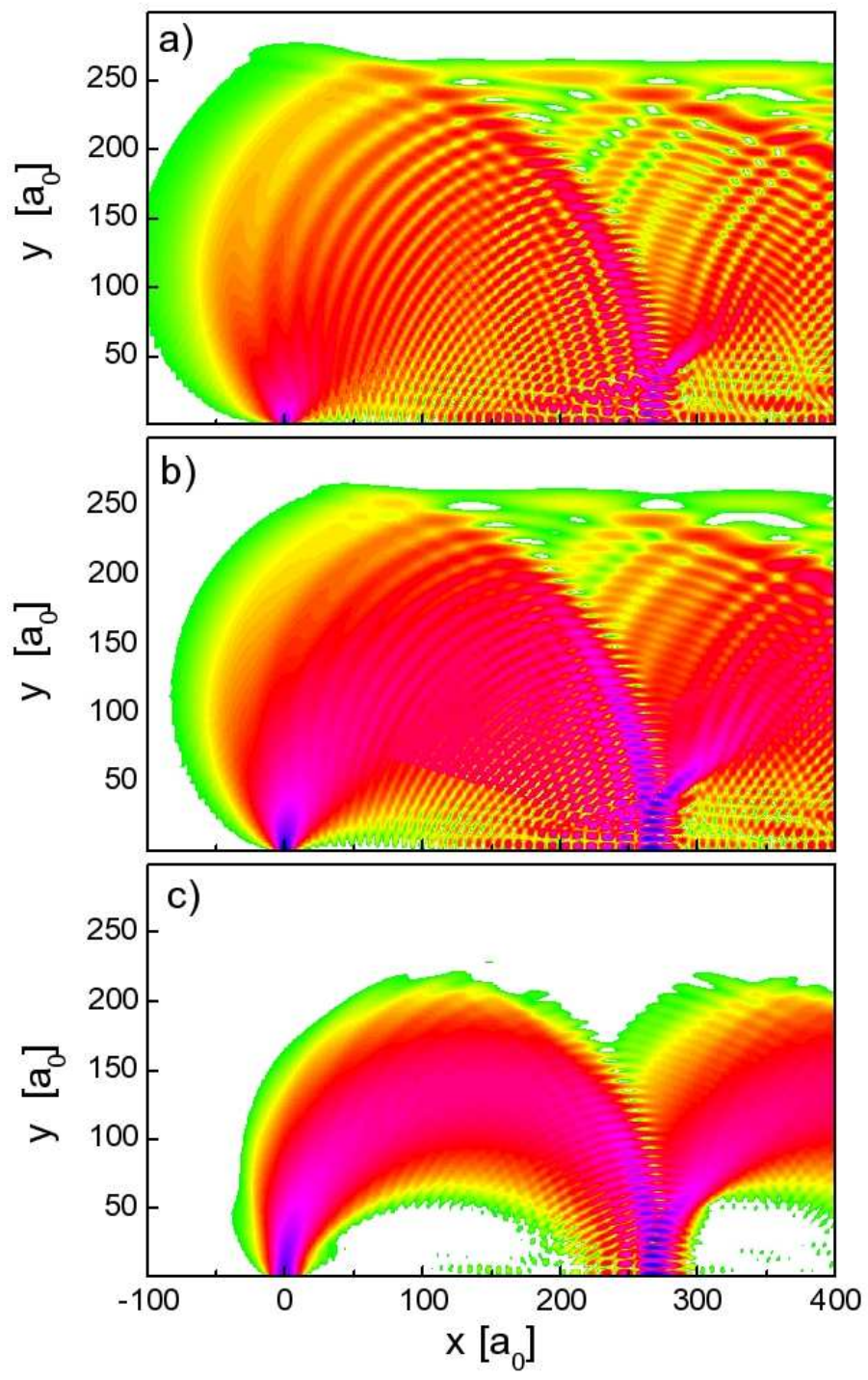

FIG. 3: (color online) Density plot of $G_{1n}$ for $B_z = 100\,mT$, $\alpha = 0$, $E_\mathrm{F} = 6\,meV$ ($s \simeq 35$) and: a) $N_0 = 5$, b) $N_0 = 13$, and c) $N_0 = 25$. Notice how the propagator peaks around the classical orbit as the width of the contact increases. According to Eq. (8), the classical cyclotron radius is $r_c \simeq 128a_0$ in good agreement with the numerical data.

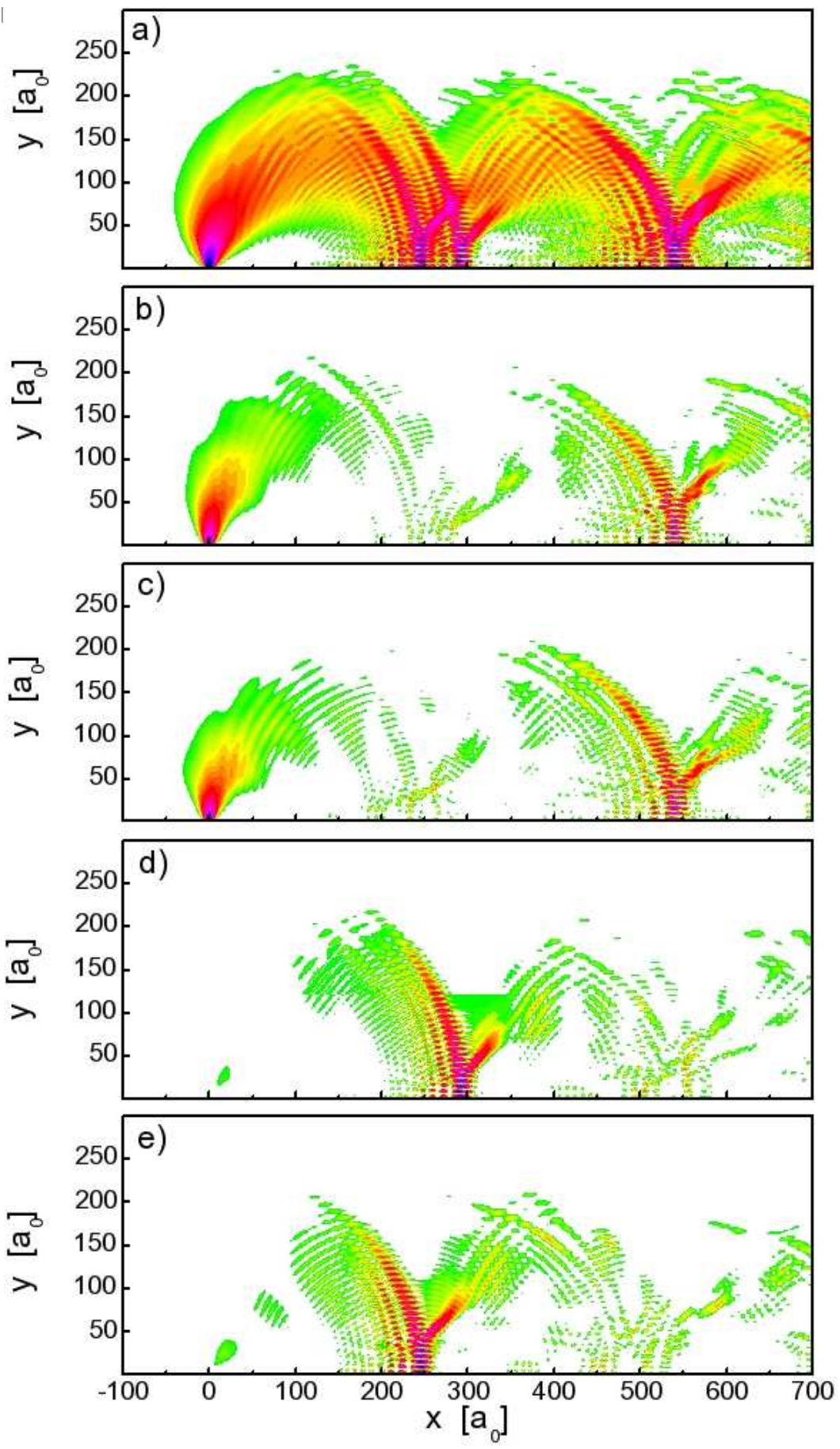

FIG. 4: (color online) Density plot of total conductance $G_{1n}$ (a) and the four spin contributions, $G_{1+,n+}$ (b), $G_{1-,n-}$ (c), $G_{1+,n-}$ (d), and $G_{1-,n+}$ (e), for $B_z = 100mT$, $\alpha = 10\,meVnm$ and $N_0 = 13$ . The presence of two cyclotron radii is apparent. Note that $2\Delta r_c \simeq 4\alpha/\hbar\omega_c = 46a_0$ in this case.

Here $L_x$ is the length of the system along the $x$-direction and $\phi_s(y-y_c)$ is the usual harmonic oscillator wavefunction with quantum number $s$ centered at $y_c = l_c^2 k_x$. The expectation value of $(y-y_c)^2$ is related to the square of the radius of the classical orbit, $r_c$. In the large $s$ limit,

$$r_c^2 = 2\langle\Psi_s^\pm|(y-y_c)^2|\Psi_s^\pm\rangle \simeq 2l_c^2 s \tag{8}$$

According to Eq. (6), within a small energy window around $E_\mathrm{F}$, the eigenstates of the two branches have different index $s$ and thus different classical orbit radii. The difference $\Delta r_c$ between the two radii is given by $\Delta r_c \simeq 2\alpha/\hbar\omega_c$. It is worth emphasizing that the peak's splitting, both in space ($\Delta r_c$) and magnetic field ($\Delta B_z$), is independent of energy. That is, it only depends on the heterostructure properties.

To illustrate the effect of two different cyclotron radii, it is convenient to remove contact 2 and calculate the conductance $G_{1n}$ from contact 1 to an arbitrary site $n$ of the semi-infinite 2DEG. This conductance gives the transmittance for an electron injected at contact 1 and collected at an arbitrary point $n$. In figure 3 we present a density plot of $G_{1n}$ for a system with $\alpha = 0$ and different contact widths. For narrow contacts (figures 3a and 3b), the caustic curves discussed by van Houten *et al.* [1] are clearly observed. They are responsible for the interference pattern observed around each peak in figure 2. As the contact width increases, the momentum $p_x$ of the incoming electron is better defined and $G_{1n}$ peaks around a single classical trajectory. Notice that the cyclotron radius observed in figure 3 is in good agreement with the value $r_c \simeq 128a_0$ obtained from Eq. (8).

The results for a system with spin-orbit coupling ($\alpha = 10\,meVnm$) and an intermediate contact width ($N_0 = 13$) are shown in figure 4. The total conductance $G_{1n}$ (top panel) clearly shows a structure corresponding to two differ-ent cyclotron radii. Figures 4b to 4e present the conductance $G_{1\sigma,n\sigma'}$ for injected and collected electrons having different spin polarizations along the $x$-axis. The four components are large in a region of space close to the classical trajectory. Near the contact, $G_{1+,n+}$ and $G_{1-,n-}$ are large while the spin-flip components $G_{1+,n-}$ and $G_{1-,n+}$ are small. As the coordinate of site $n$ departs from the contact, the former decrease while the spin-flip components increase. This clearly shows how the spin rotates along the semi-classical trajectories as the momentum rotates. At the first focusing point on the surface the conductance is dominated by the spin-flip contributions. Each contribution $G_{1+,n-}$ and $G_{1-,n+}$ presents its own cyclotron radius. The difference between the two radii is in good agreement with the argument presented above, $\Delta r_c \simeq 2\alpha/\hbar\omega_c = 23a_0$. At the second focusing point on the surface, the conductance is dominated once again by the diagonal contributions $G_{1+,n+}$ and $G_{1-,n-}$. The reason for this is that the reflection at the surface preserves the spin direction and therefore mixes the two bulk branches.[18] Thus, a state with a large orbital radius is reflected onto a state with a small radius and vice versa. In fact, as we show in figure 5, for large Rashba coupling the odd focusing points at the surface are split while the even ones are not. It is possible then to select one of the two semiclassical orbits by selecting the spin polarization of the injected electron. This is illustrated in figure 5. Conversely, an unpolarized incident beam will be split at the first focusing point in two spin-polarized beams.

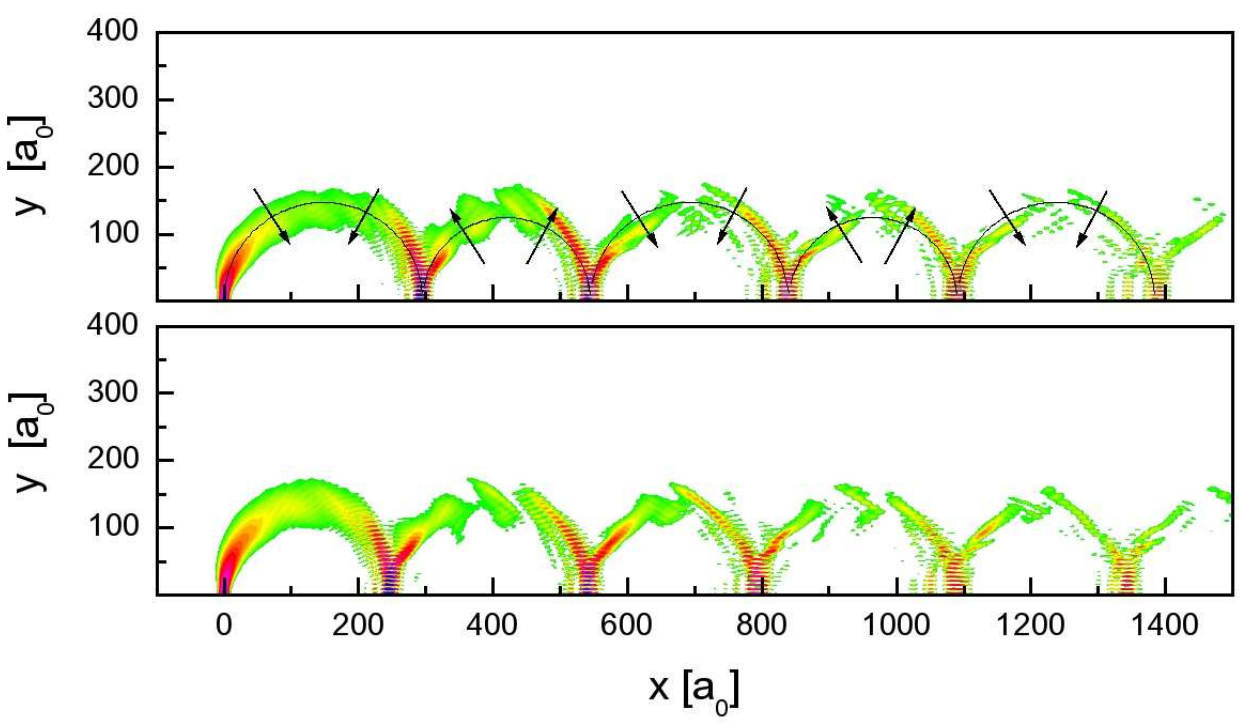

FIG. 5: (color online) Density plot of $G_{1+,n} = G_{1+,n+} + G_{1+,n-}$ (top panel) and $G_{1-,n} = G_{1-,n+} + G_{1-,n-}$ (low panel) for the same parameters as in Figure 4. The solid line in the top panel shows a semiclassical orbit with two alternating radii. The arrows represent the dominant spin projection.

In summary, we have analyzed the transverse focusing of electrons in 2DEG with Rashba spin-orbit coupling. We showed that in the weak magnetic field regime and for a given energy, the two branches of states have different cyclotron radii. This effect generates a splitting of the first focusing peak. Notably, because the sample edge mixes the two branches, the second focusing peak does not split. Higher order peaks become broader and a splitting of the odd peaks is observed only for large Rashba coupling. For parameters corresponding to GaAs/AlGaAs with a spin orbit parameter $\alpha \simeq 0.5 - 1\, meVnm$, the splitting of the first peak is small and probably hard to be observed. However, in systems like InSb/InAlSb, where $\alpha \simeq 10\, meVnm$, the effect of the spin-orbit coupling in transverse focusing experiments should be clearly observed.

*Note added in proof.* After submission of this manuscript, the splitting of the first focusing peak in a two dimensional *hole* gas in GaAs was reported in Ref. [25].

We appreciate helpful discussions with B. Alascio. This work was partially supported by ANPCyT Grant N. 99 3-6343, CNRS-PICS Collaboration Program between France and Argentina, and Fundación Antorchas, Grant 14169/21. GU acknowledge financial support from CONICET.